\documentclass[aip,jcp,reprint]{revtex4-1}

\usepackage[utf8]{inputenc}
\usepackage[T1]{fontenc}
\usepackage{amsthm, amsmath, amssymb}
\usepackage{physics}
\usepackage{graphicx}
\graphicspath{{figures/}}
\usepackage[colorlinks=true,allcolors=blue,breaklinks=true]{hyperref}

\begin{document}

\title{Two-Mode Floquet Fewest Switches Surface Hopping for Nonadiabatic Dynamics Driven by Two-Frequency Laser Fields}

\author{Jiayue Han}
\affiliation{Department of Chemistry, School of Science and Research Center for Industries of the Future, Westlake University, Hangzhou, Zhejiang 310030, China}

\author{Vahid Mosallanejad}
\affiliation{Department of Chemistry, School of Science and Research Center for Industries of the Future, Westlake University, Hangzhou, Zhejiang 310030, China}
\affiliation{Institute of Natural Sciences, Westlake Institute for Advanced Study, Hangzhou, Zhejiang 310024, China}

\author{Ruihao Bi}
\affiliation{Department of Chemistry, School of Science and Research Center for Industries of the Future, Westlake University, Hangzhou, Zhejiang 310030, China}

\author{Wenjie Dou}
\email{douwenjie@westlake.edu.cn}
\affiliation{Department of Chemistry, School of Science and Research Center for Industries of the Future, Westlake University, Hangzhou, Zhejiang 310030, China}
\affiliation{Institute of Natural Sciences, Westlake Institute for Advanced Study, Hangzhou, Zhejiang 310024, China}
\affiliation{Key Laboratory for Quantum Materials of Zhejiang Province, Department of Physics, School of Science and Research Center for Industries of the Future, Westlake University, Hangzhou, Zhejiang 310030, China}

\date{\today}

\begin{abstract}
Two-frequency (two-color) laser fields provide a powerful and flexible means for steering molecular dynamics. However, quantitatively reliable and scalable theoretical tools for simulating laser-driven nonadiabatic processes under such fields remain limited. Here, we develop a two-mode Floquet fewest switches surface hopping (two-mode F-FSSH) approach for two-frequency driving within a mixed quantum--classical framework. We validate the algorithm on three driven one-dimensional two-state models: a Rabi model and two avoided-crossing scattering models. The electronic and nuclear dynamics are benchmarked against numerically exact results from split-operator calculations, showing good agreement across a broad range of field parameters and initial conditions. These results establish two-mode F-FSSH as a practical framework for simulating and designing two-frequency control protocols and motivate extensions to more realistic experimental settings.
\end{abstract}

\maketitle

\section{Introduction}

Strong laser fields offer a highly tunable route to manipulate molecular dynamics far from equilibrium, underpinning a wide range of phenomena in physical chemistry and materials science.\cite{Zhang2019non,Weight2023theory,Wu2025light} Representative examples include ultrafast spectroscopy,\cite{Maiuri2019ultrafast,Thoss2007correlated} coherent control,\cite{Bandrauk1981photodissociation,Corrales2014control,Corrales2017strong} photocatalysis,\cite{Maeda2006photocatalyst,Zhao2020two} and plasmonic chemistry.\cite{Zhang2017surface,Sarkar2025plasmon} Among various driving protocols, periodic two-frequency fields (often termed two-color fields in experiments) have attracted considerable attention\cite{Tak_San_Ho1984semiclassical,Su1996simple,Il_in2014theory}. Their structured spectra and temporal modulation provide additional control parameters, including independently tunable amplitudes, relative phase, polarization, and frequency detuning between the two components. This expanded controllability has made periodic two-frequency schemes valuable across multiple frontiers. In nonlinear optics, two-frequency fields are widely used to tailor high-order harmonic generation (HHG) and related frequency-conversion processes.\cite{Lan2010wavelength,Raab2024highly} Two-frequency excitation also plays a central role in ultrafast spectroscopy, enabling broadly tunable pump--probe configurations and multidimensional spectroscopies.\cite{Manzoni2006two,Myers2008two} In precision diagnostics and metrology, interferometric techniques (e.g., two-frequency interferometry and second-harmonic dispersion interferometry) allow robust retrieval of plasma- and density-related observables.\cite{Sagisaka2006development,Cheng2012dual,Chavez2023measurement,Qin2025a} Moreover, in ultrafast imaging of rapid transients, two-frequency illumination can improve reconstruction fidelity and enhance image contrast.\cite{Hayasaki2017two}

These rapid experimental advances in periodic two-frequency laser techniques have stimulated theoretical efforts to elucidate the underlying dynamical mechanisms and to enable rational control strategies. A central theoretical challenge is that the Born--Oppenheimer (BO) approximation breaks down under strong external driving, where field-induced electronic transitions become intertwined with nuclear motion through nonadiabatic couplings.\cite{Tully2012perspective,Subotnik2016understanding} A direct and fully quantum description is obtained by solving the full time-dependent Schr\"odinger equation (TDSE), for example via quantum wavepacket dynamics\cite{Deumens1994time,Suzuki2014electronic} or the multiconfiguration time-dependent Hartree (MCTDH) method.\cite{Beck2000the,Wang2003multilayer} However, TDSE-based approaches typically incur a steep computational cost that grows rapidly with the number of nuclear degrees of freedom, which limits their routine applications to small systems. This consideration motivates mixed quantum--classical dynamics, most notably Tully's fewest switches surface hopping (FSSH)\cite{Tully1990molecular} and its variants, which often offer a practical balance between numerical accuracy and computational efficiency.\cite{Crespo_Otero2018recent}

To incorporate external driving into FSSH, two major strategies have been pursued. One is instantaneous adiabatic fewest switches surface hopping (IA-FSSH), where the explicitly time-dependent electronic Hamiltonian is diagonalized on-the-fly to obtain instantaneous adiabatic states and potential energy surfaces (PESs). This method has been implemented in several practical schemes, including the surface hopping including arbitrary couplings (SHARC) package\cite{Richter2011sharc,Mai2015a,Mai2018nonadiabatic} and related field-induced surface hopping (FISH) approaches.\cite{Mitric2011field,Lisinetskaya2011simulation} While conceptually straightforward, IA-FSSH can face limitations under strong or resonant driving. First, the instantaneous adiabatic PESs may oscillate rapidly, leading to intricate electronic phase relations and repeated wavepacket splitting and remerging. Such dynamics can introduce recoherence events that are not naturally captured by trajectory-based surface hopping.\cite{Subotnik2011fewest,Subotnik2011a,Landry2012how,Subotnik2013can} A second limitation is energy consistency. In standard (field-free) FSSH, nuclear momenta are typically rescaled during a hopping event to conserve the total energy. Under external driving, however, energy is not conserved for the molecular subsystem because it can be continuously exchanged with the field.\cite{Zhou2020nonadiabatic,Zhou2023nonadiabatic} To date, there is no universally accepted prescription for handling energy consistency and momentum rescaling in IA-FSSH under strong time-dependent driving. In addition, accurate evaluation of time derivative couplings and consistent phase tracking of instantaneous adiabatic states can be challenging in practice.\cite{Shenvi2011phase,Mai2015a,Zhou2019a}

Motivated by these challenges, an alternative strategy is to combine fewest switches surface hopping with Floquet theory by recasting a periodic time-dependent Hamiltonian into a time-independent one.\cite{Floquet1883sur,Shirley1965solution} This Floquet fewest switches surface hopping (F-FSSH) approach has been extensively explored for periodically driven systems.\cite{Bajo2012mixed,Zhou2020nonadiabatic,Zhou2023nonadiabatic,Wang2023nonadiabatic_FSH,Wang2024nonadiabatic,Wang2023electron} Within this framework, nuclear trajectories propagate on Floquet quasi-energy surfaces and undergo hops between them. Thus far, F-FSSH has been developed primarily for single-frequency driving cases. A general F-FSSH approach for two-frequency driving has not yet been established.

Here, we develop a two-mode Floquet fewest switches surface hopping (two-mode F-FSSH) approach for two-frequency laser-driven nonadiabatic dynamics. In Sec.~\ref{sec:TheoryandMethodology}, we present the theoretical formulation and the corresponding two-mode F-FSSH algorithm. In Sec.~\ref{sec:SimulationDetails}, we introduce the model Hamiltonians and provide simulation details for both the exact and two-mode F-FSSH calculations. In Sec.~\ref{sec:ResultsandDiscussion}, we benchmark the two-mode F-FSSH algorithm by comparing electronic and nuclear dynamics with numerically exact calculations. Finally, we summarize the main findings and outline future directions in Sec.~\ref{sec:Conclusions}.

\section{Theory and Methodology}
\label{sec:TheoryandMethodology}

\subsection{Molecular Hamiltonian Driven by Strong Laser Fields}

We consider laser-driven nonadiabatic molecular dynamics, where the total Hamiltonian is written as
\begin{equation}
\hat{H}(t) = \hat{T}_{\mathrm{n}} + \hat{H}_{\mathrm{el}}(\hat{\mathbf{R}}) + \hat{V}_{\mathrm{int}}(t).
\label{eq:H_total}
\end{equation}
Here $\hat{\mathbf{R}}$ denotes the nuclear coordinate operator, $\hat{T}_{\mathrm{n}}$ is the nuclear kinetic-energy operator, and $\hat{H}_{\mathrm{el}}(\hat{\mathbf{R}})$ is the field-free electronic Hamiltonian. Throughout this work, we employ the electric-dipole approximation for the laser--molecule interaction,
\begin{equation}
\hat{V}_{\mathrm{int}}(t) = - \hat{\boldsymbol{\mu}}(\hat{\mathbf{R}}) \cdot \mathbf{E}(t),
\label{eq:V_int}
\end{equation}
where $\hat{\boldsymbol{\mu}}(\hat{\mathbf{R}})$ is the molecular dipole operator and $\mathbf{E}(t)$ is the applied electric field. We consider a two-frequency field composed of two components with frequencies $\omega_1$ and $\omega_2$ along the unit vector $\mathbf{e}$,
\begin{equation}
\mathbf{E}(t) = \mathbf{e}\Big[ E_1 \cos\!\big(\omega_1 t + \phi_1\big)
+ E_2 \cos\!\big(\omega_2 t + \phi_2\big)\Big],
\label{eq:E_twofreq_general}
\end{equation}
where $E_{1,2}$ are the field amplitudes and $\phi_{1,2}$ are the carrier phases.

\subsection{Two-Mode Floquet Formalism}

We now briefly review the general two-mode Floquet formalism.\cite{Mosallanejad2025two} For a closed system, we begin with the time-independent Floquet Liouville--von Neumann equation in the extended space,
\begin{equation}
\frac{d\hat{\rho}^{F}(t)}{dt}
=
-\frac{i}{\hbar}\,\commutator{\hat{H}^{F}}{\hat{\rho}^{F}(t)},
\label{eq:FLvN}
\end{equation}
where $\hat{H}^{F}$ is the two-mode Floquet Hamiltonian and $\hat{\rho}^{F}(t)$ is the corresponding two-mode Floquet density operator.

To construct $\hat{H}^{F}$, we rewrite the time-dependent Hamiltonian in a two-time form $\hat{H}(t_1,t_2)$ by reindexing the oscillatory terms with frequencies $\omega_1$ and $\omega_2$ using independent time variables $t_1$ and $t_2$, respectively. The original Hamiltonian is recovered on the diagonal $t_1=t_2=t$, namely $\hat{H}(t)\equiv \hat{H}(t,t)$. We then expand $\hat{H}(t_1,t_2)$ in a two-dimensional Fourier series,
\begin{equation}
\hat{H}(t_1,t_2)
=
\sum_{m,n\in\mathbb{Z}}
\hat{H}^{mn}\,
e^{i n\omega_1 t_1}\,
e^{i m\omega_2 t_2},
\label{eq:H_two_mode_fourier}
\end{equation}
where the coefficient operators are given by
\begin{equation}
\hat{H}^{mn}
=
\frac{1}{T_1 T_2}
\int_{0}^{T_2}\! dt_2
\int_{0}^{T_1}\! dt_1\,
\hat{H}(t_1,t_2)\,
e^{-i n\omega_1 t_1}\,
e^{-i m\omega_2 t_2},
\label{eq:Hmn_def}
\end{equation}
with $T_1=2\pi/\omega_1$ and $T_2=2\pi/\omega_2$.

Next, we introduce the two-mode Fourier basis $\{\ket{m,n}\}$ for the extended space, where $n$ labels the Fourier index associated with $\omega_1$ and $m$ labels that associated with $\omega_2$.
In this basis, the two families of ladder operators $\hat{L}^{\prime}_{k}$ and $\hat{L}^{\prime\prime}_{k}$ and the corresponding number operators $\hat{N}^{\prime}$ and $\hat{N}^{\prime\prime}$ act as
\begin{align}
\hat{L}^{\prime}_{k}\ket{m,n} &= \ket{m,n+k},
&
\hat{L}^{\prime\prime}_{k}\ket{m,n} &= \ket{m+k,n}, \nonumber\\
\hat{N}^{\prime}\ket{m,n} &= n\ket{m,n},
&
\hat{N}^{\prime\prime}\ket{m,n} &= m\ket{m,n}, \nonumber\\
\commutator{\hat{N}^{\prime}}{\hat{L}^{\prime}_{k}} &= k\,\hat{L}^{\prime}_{k},
&
\commutator{\hat{N}^{\prime\prime}}{\hat{L}^{\prime\prime}_{k}} &= k\,\hat{L}^{\prime\prime}_{k}, \nonumber\\
\commutator{\hat{L}^{\prime}_{n}}{\hat{L}^{\prime\prime}_{k}} &= 0,
&
\commutator{\hat{N}^{\prime}}{\hat{N}^{\prime\prime}} &= 0.
\label{eq:two_mode_actions}
\end{align}

With the Fourier components $\hat{H}^{mn}$ and the operators in Eq.~\eqref{eq:two_mode_actions}, the time-independent two-mode Floquet Hamiltonian $\hat{H}^{F}$ is assembled as
\begin{equation}
\hat{H}^{F}
=
\sum_{m,n}
\hat{L}^{\prime\prime}_{m}\hat{L}^{\prime}_{n}
\otimes
\hat{H}^{mn}
+
\hat{N}^{\prime}\otimes \hat{I}\,\hbar\omega_1
+
\hat{N}^{\prime\prime}\otimes \hat{I}\,\hbar\omega_2,
\label{eq:H_F_two_mode}
\end{equation}
where $\hat{I}$ denotes the identity operator in the electronic Hilbert space. Similarly, the two-mode Floquet density operator is defined as
\begin{equation}
\hat{\rho}^{F}(t)
=
\sum_{m,n}
\hat{L}^{\prime\prime}_{m}\hat{L}^{\prime}_{n}
\otimes
\hat{\rho}^{mn}(t),
\label{eq:rho_F_two_mode}
\end{equation}
where $\hat{\rho}^{mn}(t)$ are the corresponding coefficient operators.

Physical observables are defined in the electronic Hilbert space. Accordingly, the physical density operator $\hat{\rho}(t)$ is obtained by projecting the Floquet density operator $\hat{\rho}^{F}(t)$ back to the original Hilbert space,
\begin{equation}
\hat{\rho}(t)
=
\sum_{m,n}
\bra{m,n}\hat{\rho}^{F}(t)\ket{0,0}\,
e^{i n\omega_1 t}\,e^{i m\omega_2 t}.
\label{eq:projection_to_Hilbert}
\end{equation}
Here, we select the reference replica $\ket{0,0}$, and the phase factors restore the explicit time dependence in the physical representation.

\subsection{Two-Mode Floquet Fewest Switches Surface Hopping}

By applying a partial Wigner transform\cite{_mre1967wigner,Kapral1999mixed} to Eq.~\eqref{eq:FLvN} with respect to the nuclear degrees of freedom, we obtain the Floquet quantum--classical Liouville equation (F-QCLE):\cite{Mosallanejad2023floquet}
\begin{equation}
\begin{aligned}
\pdv{\hat{\rho}^{F}_{\mathrm{W}}(t)}{t}
&=
-\frac{i}{\hbar}\,\commutator{\hat{H}^{F}_{\mathrm{el}}}{\hat{\rho}^{F}_{\mathrm{W}}(t)}
-\sum_{\alpha}\frac{P_{\alpha}}{M_{\alpha}}\,\pdv{\hat{\rho}^{F}_{\mathrm{W}}(t)}{R_{\alpha}} \\
&\quad
+\frac{1}{2}\sum_{\alpha}\anticommutator{
\pdv{\hat{H}^{F}_{\mathrm{el}}}{R_{\alpha}}
}{
\pdv{\hat{\rho}^{F}_{\mathrm{W}}(t)}{P_{\alpha}}
}.
\end{aligned}
\label{eq:fqcle}
\end{equation}
Here $\hat{\rho}^{F}_{\mathrm{W}}(t)$ denotes the partial Wigner transforms of $\hat{\rho}^{F}(t)$, such that the nuclear operators are replaced by the corresponding classical phase-space variables $\mathbf{R}=\{R_{\alpha}\}$ and $\mathbf{P}=\{P_{\alpha}\}$. The index $\alpha$ labels the classical nuclear degrees of freedom, and $\anticommutator{\hat{A}}{\hat{B}}\equiv \hat{A}\hat{B}+\hat{B}\hat{A}$ defines the anticommutator.

Eq.~\eqref{eq:fqcle} provides a mixed quantum--classical description, which naturally motivates a trajectory-based surface hopping implementation in the two-mode Floquet representation. For each nuclear geometry $\mathbf{R}$, we construct a Floquet electronic Hamiltonian $\hat{H}^{F}_{\mathrm{el}}(\mathbf{R})$ and diagonalize it to obtain the Floquet adiabatic states and quasi-energies,
\begin{equation}
\hat{H}^{F}_{\mathrm{el}}(\mathbf{R})\ket{\Phi^{F}_{j}(\mathbf{R})}
=
E_{j}(\mathbf{R})\ket{\Phi^{F}_{j}(\mathbf{R})}.
\label{eq:HF_eig}
\end{equation}

Each nuclear trajectory evolves on the active Floquet quasi-energy surface $E_{j}(\mathbf{R})$ associated with the active Floquet state $j$. The nuclear degrees of freedom are propagated classically according to Newton's equations of motion,
\begin{equation}
\dot{R}_{\alpha}=\frac{P_{\alpha}}{M_{\alpha}},
\qquad
\dot{P}_{\alpha}=-\pdv{E_{j}(\mathbf{R})}{R_{\alpha}}.
\label{eq:newton_F}
\end{equation}
The Floquet electronic density matrix is propagated either (i) in the diabatic Floquet basis by directly integrating Eq.~\eqref{eq:FLvN}, or (ii) in the adiabatic Floquet basis obtained by diagonalizing $\hat{H}^{F}_{\mathrm{el}}(\mathbf{R})$. In the adiabatic representation, we define Floquet adiabatic density matrix elements as
$\rho^{F}_{jk}(t)=\mel{\Phi^{F}_{j}(\mathbf{R}(t))}{\hat{\rho}^{F}(t)}{\Phi^{F}_{k}(\mathbf{R}(t))}$. The equation of motion is given by
\begin{equation}
i\hbar\,\dot{\rho}^{F}_{jk}
=
\sum_{l}\Bigl[
\rho^{F}_{lk}\Bigl(E_{j}\delta_{jl}-i\hbar\,\dot{\mathbf{R}}\cdot \mathbf{d}^{F}_{jl}\Bigr)
-
\rho^{F}_{jl}\Bigl(E_{l}\delta_{lk}-i\hbar\,\dot{\mathbf{R}}\cdot \mathbf{d}^{F}_{lk}\Bigr)
\Bigr].
\label{eq:rhoF_adiabatic_eom}
\end{equation}
where $\mathbf{d}^{F}_{jk}(\mathbf{R})=\mel{\Phi^{F}_{j}(\mathbf{R})}{\nabla_{\mathbf{R}}}{\Phi^{F}_{k}(\mathbf{R})}$ denotes the Floquet derivative coupling between adiabatic Floquet states $j$ and $k$.

Following Tully's fewest-switches prescription,\cite{Tully1990molecular} the hopping probability from the active surface $j$ to the target surface $k$ within one time step $\Delta t$ is
\begin{equation}
g_{j\rightarrow k}
=
\max\!\left[
0,\,
-2\,\Re\!\left\{
\dot{\mathbf{R}}\cdot \mathbf{d}^{F}_{kj}\,
\frac{\rho^{F}_{jk}}{\rho^{F}_{jj}}
\right\}
\right]\Delta t.
\label{eq:hop_prob_F}
\end{equation}
Upon a successful hop $j\to k$, the nuclear momentum is rescaled to conserve the total energy,
\begin{equation}
\sum_{\alpha}\frac{P_{\alpha}^{2}}{2M_{\alpha}}+E_{j}(\mathbf{R})
=
\sum_{\alpha}\frac{P_{\alpha}^{\prime\,2}}{2M_{\alpha}}+E_{k}(\mathbf{R}).
\label{eq:energy_conservation_F}
\end{equation}
where $P_{\alpha}^{\prime}$ denotes the post-hop momentum. The momentum rescaling is performed along the direction of $\Re\!\left(\mathbf{d}^{F}_{jk}\right)$,
\begin{equation}
\mathbf{n}^{F}_{jk}
=
\frac{\Re\!\left(\mathbf{d}^{F}_{jk}\right)}
{\left|\Re\!\left(\mathbf{d}^{F}_{jk}\right)\right|}.
\label{eq:rescale_dir_F}
\end{equation}
If Eq.~\eqref{eq:energy_conservation_F} admits no real solution for $P_{\alpha}^{\prime}$ when rescaling, the hop is rejected.

In principle, the two-mode Floquet Hamiltonian $\hat{H}^{F}(\mathbf{R})$ acts on an infinite-dimensional extended space spanned by the Fourier indices $(m,n)\in\mathbb{Z}^2$, which renders two-mode F-FSSH computationally impractical. We therefore truncate the two Fourier spaces by retaining $n\in[-N_1,N_1]$ for the $\omega_1$ component and $m\in[-N_2,N_2]$ for the $\omega_2$ component. The resulting Floquet Hamiltonian dimension is $D_F = D_{\mathrm{e}}(2N_1+1)(2N_2+1)$, where $D_{\mathrm{e}}$ is the dimension of the electronic Hilbert space. In practice, $N_1$ and $N_2$ are increased until the target physical observables (e.g., electronic populations) are converged within the desired accuracy.

\section{Simulation Details}
\label{sec:SimulationDetails}

\subsection{Model Hamiltonians}

In this subsection, we introduce three driven one-dimensional two-state model Hamiltonians that serve as test systems for assessing the accuracy and robustness of surface hopping algorithms. Unless otherwise noted, all quantities are reported in atomic units (a.u.).

\begin{figure*}[t]
\centering
\includegraphics[width=\textwidth]{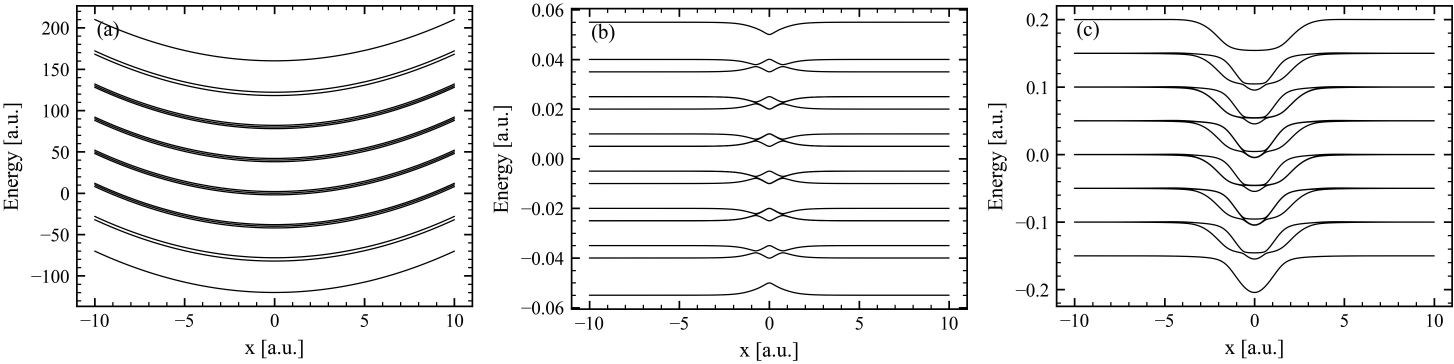}
\caption{Adiabatic Floquet quasi-energy surfaces for the model systems considered in this work. For visualization, the two-mode Floquet Hamiltonian is truncated with $N_1=N_2=1$: (a) driven Rabi model with $E_1=E_2=4.0$, $\omega_1=40.0$, and $\omega_2=2\omega_1$; (b) simple avoided-crossing model with $E_1=E_2=0.3$, $\omega_1=0.015$, and $\omega_2=2\omega_1$; and (c) dual avoided-crossing model with $E_1=E_2=0.3$, $\omega_1=0.05$, and $\omega_2=2\omega_1$. The dense replica structure in the avoided-crossing models gives rise to numerous trivial crossings that can cause numerical instabilities.}
\label{fig:floquet_surfaces}
\end{figure*}

\subsubsection{Rabi Model}

As a first model system, we consider a driven Rabi model with two vertically displaced one-dimensional harmonic potentials.\cite{Richter2011sharc} In the diabatic electronic basis $\{\ket{1},\ket{2}\}$, the field-free electronic Hamiltonian reads
\begin{equation}
\hat{H}_e(x)
=
\begin{pmatrix}
\frac{1}{2}kx^2 & 0 \\
0 & \frac{1}{2}kx^2+\Delta
\end{pmatrix},
\label{eq:rabi_H0}
\end{equation}
where $x$ is the nuclear coordinate, $k$ is the harmonic force constant, and $\Delta$ is the vertical energy gap between the two diabatic potentials. In this work, we set $k=1$ and $\Delta=40$.

The transition dipole operator is assumed to be purely off-diagonal and independent of the nuclear coordinate,
\begin{equation}
\hat{\mu}
=
\begin{pmatrix}
0 & 1 \\
1 & 0
\end{pmatrix}.
\label{eq:rabi_mu}
\end{equation}
For this model, the applied two-frequency field is
\begin{equation}
E(t)=E_1\sin\!\big(\omega_1 t\big)+E_2\sin\!\big(\omega_2 t\big),
\label{eq:rabi_twofreq_field}
\end{equation}
which corresponds to Eq.~\eqref{eq:E_twofreq_general} with carrier phases $\phi_1=\phi_2=-\pi/2$.

Fig.~\ref{fig:floquet_surfaces}(a) illustrates representative Floquet quasi-energy surfaces for the driven Rabi model under the driving parameters specified in the caption.

\subsubsection{Avoided-Crossing Scattering Models}

To benchmark two-mode F-FSSH on field-controlled nonadiabatic scattering, we consider two standard one-dimensional avoided-crossing models proposed by Tully.\cite{Tully1990molecular} In the diabatic basis $\{\ket{1},\ket{2}\}$, the explicitly time-dependent electronic Hamiltonian is written as
\begin{equation}
\hat{H}_e(x,t)=
\begin{pmatrix}
V_{11}(x) & W(x,t) \\
W(x,t) & V_{22}(x)
\end{pmatrix}.
\label{eq:ac_general_He}
\end{equation}
In both models, the diabatic coupling is modulated by a two-frequency field,
\begin{equation}
W(x,t)=W_0(x)\Big[1+E_1\cos(\omega_1 t)+E_2\cos(\omega_2 t)\Big],
\label{eq:ac_general_Wxt_common}
\end{equation}
so that the driving primarily affects the nonadiabatic coupling region. The field-free limit is recovered by setting $E_1=E_2=0$, in which case the time dependence vanishes and the diabatic coupling reduces to $W_0(x)$.

\paragraph{Simple Avoided Crossing.}
For the simple avoided-crossing model, we take
\begin{equation}
V_{11}(x)=
\begin{cases}
A\!\left[1-e^{-Bx}\right], & x>0,\\
-A\!\left[1-e^{Bx}\right], & x\le 0,
\end{cases}
\label{eq:ac1_V11}
\end{equation}
and set $V_{22}(x)=-V_{11}(x)$, with
\begin{equation}
W_0(x)=C\,e^{-Dx^{2}}.
\label{eq:ac1_W0}
\end{equation}
Here we use $A=0.01$, $B=1.6$, $C=0.005$, and $D=1.0$.

\paragraph{Dual Avoided Crossing.}
For the dual avoided-crossing model, we take
\begin{align}
V_{11}(x) &= 0, \label{eq:ac2_V11}\\
V_{22}(x) &= -A\,e^{-Bx^{2}}+E_0, \label{eq:ac2_V22}\\
W_0(x) &= C\,e^{-Dx^{2}}, \label{eq:ac2_W0}
\end{align}
which generates two well-separated avoided crossings as the wavepacket traverses the interaction region. Here we use $A=0.10$, $B=0.28$, $C=0.015$, $D=0.06$, and $E_0=0.05$.

Figs.~\ref{fig:floquet_surfaces}(b,c) illustrate representative Floquet quasi-energy surfaces for the driven avoided-crossing scattering models under the driving parameters specified in the caption. Numerous trivial crossings can compromise the numerical stability of surface hopping dynamics. Practical remedies are discussed below.

\subsection{Exact Calculations and Initial Conditions}

We obtain numerically exact results using the split-operator method proposed by Kosloff and Kosloff.\cite{Kosloff1983a} The initial nuclear wavepacket is chosen as a normalized Gaussian,
\begin{equation}
\psi(x,0)=\left(\frac{1}{\pi\sigma^{2}}\right)^{1/4}
\exp\!\left[-\frac{(x-x_0)^2}{2\sigma^{2}}+ \frac{i}{\hbar}p_0(x-x_0)\right],
\label{eq:initial_gaussian_wp}
\end{equation}
and the initial electronic state is taken to be the ground diabatic state at $t=0$. For the driven Rabi model, we place the nuclear wavepacket at the minimum of the lower harmonic potential, with $p_0=0$ and $\sigma=1/\sqrt{2}$. For the avoided-crossing scattering models, we initialize the wavepacket centered at $x_0=-10$ (far from the interaction region) with $\sigma=p_0/20$, which adjusts the spatial width with the incident momentum.\cite{Tully1990molecular}

The initial conditions for two-mode F-FSSH trajectories are sampled from the Wigner distribution corresponding to the Gaussian wavepacket in Eq.~\eqref{eq:initial_gaussian_wp}. For a minimum-uncertainty Gaussian, the Wigner function reads
\begin{equation}
W(x,p)=\frac{1}{\pi\hbar}\exp\!\left[-\frac{(x-x_0)^2}{\sigma^2}\right]
\exp\!\left[-\frac{\sigma^2(p-p_0)^2}{\hbar^2}\right],
\label{eq:wigner_gaussian}
\end{equation}
which implies that $x$ and $p$ can be sampled independently from normal distributions with standard deviations $\Delta x=\sigma/\sqrt{2}$ and $\Delta p=\hbar/(\sqrt{2}\sigma)$, respectively. Because Floquet nonadiabatic dynamics under external driving can be sensitive to the initial nuclear distribution, we ensure that the trajectory sampling uses the same $(x_0,p_0,\sigma)$ as the corresponding wavepacket benchmarks.\cite{Zhou2020nonadiabatic}

\subsection{Implementation Details of Two-Mode F-FSSH}

To improve numerical stability and reproducibility, we employ several implementation techniques in our two-mode F-FSSH simulations. These procedures are not specific to any particular model and serve as practical guidelines for robust calculations.

\subsubsection{Trivial Crossings in the Floquet Representation}

Trivial crossings are ubiquitous in Floquet representations because the enlarged Floquet space contains many near-degenerate replicas that can intersect as functions of nuclear geometry.\cite{Zhou2019a} They become particularly problematic when the Floquet derivative couplings $\mathbf{d}^{F}_{jk}(\mathbf{R})$ are evaluated via the Hellmann--Feynman expression,
\begin{equation}
\mathbf{d}^{F}_{jk}(\mathbf{R})
=
-\frac{\mel{\Phi^{F}_{j}(\mathbf{R})}{\nabla_{\mathbf{R}}\hat{H}^{F}_{\mathrm{el}}(\mathbf{R})}{\Phi^{F}_{k}(\mathbf{R})}}{E_{j}(\mathbf{R})-E_{k}(\mathbf{R})},
\qquad (j\neq k).
\label{eq:dc_HF_F}
\end{equation}
Specifically, the small quasi-energy gaps $E_j(\mathbf{R})-E_k(\mathbf{R})$ in the denominator can spuriously amplify $\mathbf{d}^{F}_{jk}$ near exact trivial crossings, which can in turn lead to erroneous hopping probabilities and distorted electronic coherences. To obtain stable and physically meaningful dynamics in the presence of trivial crossings, we employ the following two remedies simultaneously:

\paragraph{Phase Correction.}
In numerical implementations, the Floquet adiabatic states are obtained by diagonalizing $\hat{H}^{F}(\mathbf{R})$. Mathematically, if $\ket{\Phi^{F}_{j}(\mathbf{R})}$ is an eigenvector then $e^{i\theta_{j}(\mathbf{R})}\ket{\Phi^{F}_{j}(\mathbf{R})}$ is equally valid for any real $\theta_{j}(\mathbf{R})$. This intrinsic gauge freedom leads to phase indeterminacy from one geometry to the next in repeated diagonalizations.\cite{Wang2016recent} Surface hopping in the Floquet representation is sensitive to these phases. In this work, conventional parallel transport does not provide sufficiently robust phase tracking, and we therefore employ the phase correction procedure introduced by Mai \textit{et al.}\cite{Mai2015a,Mai2018nonadiabatic} to enforce phase continuity of the Floquet adiabatic states.

\paragraph{Diabatic Evolution.}
The Floquet electronic density matrix can be propagated in either the diabatic or the adiabatic representation. In principle, the two choices are equivalent and related by a unitary transformation.\cite{Wang2014a} In practice, however, adiabatic propagation requires derivative couplings and can become numerically unstable near trivial crossings, even with phase correction. We therefore propagate the Floquet electronic density matrix in the diabatic basis while evaluating hopping events in the adiabatic basis. Although this mixed-representation strategy is not common in conventional FSSH implementations, it is essential for robust and reproducible two-mode F-FSSH simulations.

\subsubsection{Frustrated Hops}

In FSSH, an attempted hop from an active surface $j$ to a target surface $k$ may be rejected when the nuclear kinetic energy along the rescaling direction is insufficient to satisfy the energy-matching requirement. Such rejected events are referred to as frustrated hops. Properly treating frustrated hops is essential for maintaining physically reasonable nuclear momentum distributions and avoiding systematic biases in state populations. In this work, we adopt the widely used $\Delta V$ prescription introduced by Jasper and Truhlar.\cite{Jasper2003improved} Briefly, when a hop $j\to k$ is attempted but rejected, the nuclear momentum is reversed along $\mathbf{n}^{F}_{jk}$ defined in Eq.~\eqref{eq:rescale_dir_F} if
\begin{equation}
\big(\nabla_{\mathbf{R}}E_{k}\cdot \mathbf{n}^{F}_{jk}\big)
\big(\mathbf{P}\cdot \mathbf{n}^{F}_{jk}\big) > 0.
\label{eq:frustrated_hop_reverse}
\end{equation}
Otherwise, the momentum is kept unchanged and the trajectory continues evolving on the current surface $j$.

\section{Results and Discussion}
\label{sec:ResultsandDiscussion}

In this section, we report electronic and nuclear dynamics for the three model Hamiltonians introduced above. Numerically exact reference results are obtained with the split-operator method, and the approximate results are generated using our two-mode Floquet fewest switches surface hopping (two-mode F-FSSH) approach.

\subsection{Rabi Model}

\begin{figure*}[t]
\centering
\includegraphics[width=\textwidth]{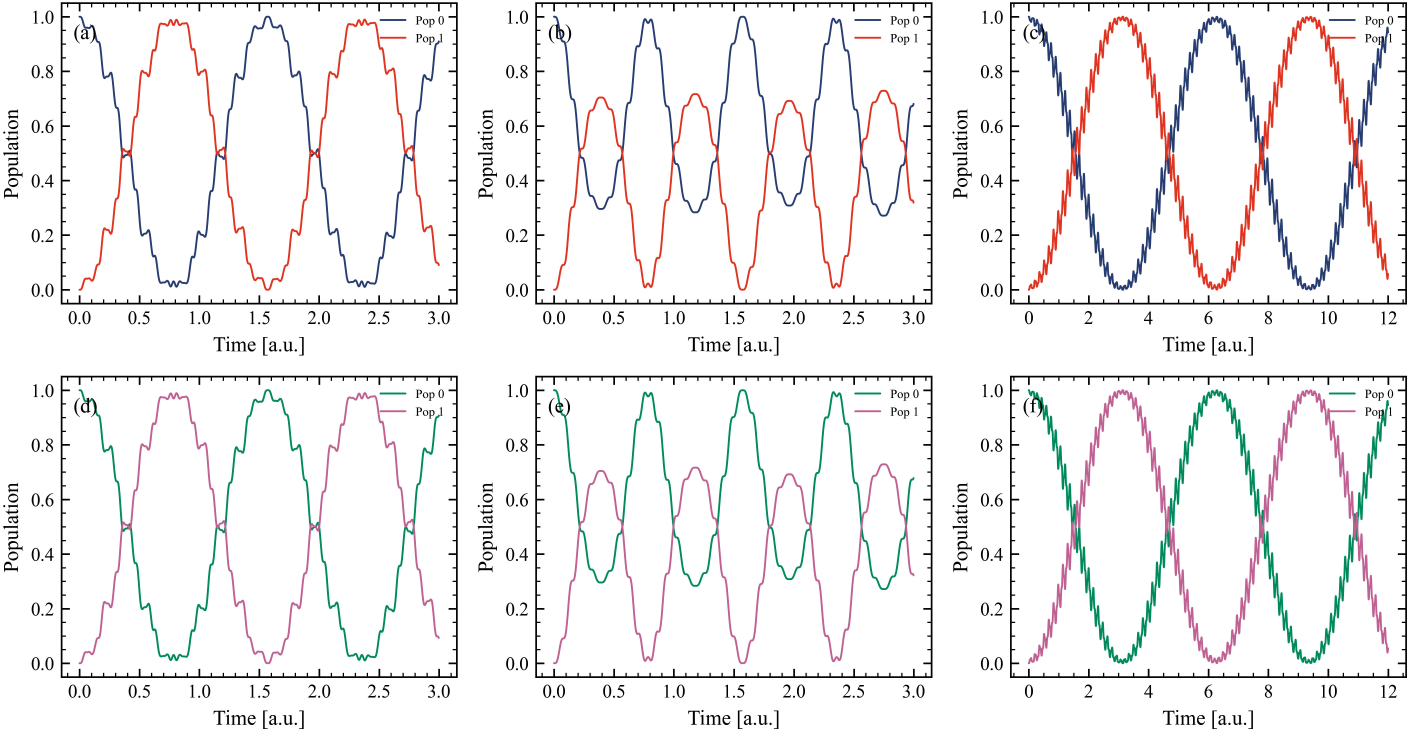}
\caption{Diabatic populations for the driven Rabi model under representative two-frequency driving. Top row: numerically exact split-operator results. Bottom row: two-mode F-FSSH results using 100 trajectories with $\Delta t=0.002$. Left column: $E_1=4$, $\omega_1=40$, $E_2=4$, $\omega_2=80$. Middle column: $E_1=4$, $\omega_1=36$, $E_2=4$, $\omega_2=44$. Right column: $E_1=1$, $\omega_1=40$, $E_2=4$, $\omega_2=80$. Overall, two-mode F-FSSH quantitatively reproduces the numerically exact split-operator predictions for the electronic population dynamics across these representative two-frequency driving conditions.}
\label{fig:rabi}
\end{figure*}

We first test two-mode F-FSSH on the driven Rabi model. Fig.~\ref{fig:rabi} compares diabatic populations obtained from numerically exact split-operator propagation (top row) and from two-mode F-FSSH (bottom row). For the two-mode F-FSSH results, we use an ensemble of 100 trajectories with a time step $\Delta t=0.002$. The two sets of curves are in excellent agreement in all cases. This agreement is expected for the present model: the field-free Hamiltonian is strictly diagonal [Eq.~\eqref{eq:rabi_H0}], and thus there are no hops due to nonadiabatic transitions. The population dynamics are driven entirely by the light--matter interaction $\hat{V}_{\mathrm{int}}(t)=-\hat{\mu}E(t)$, which induces coherent and continuous population transfer between the diabatic states.

The three columns illustrate representative combinations of driving frequencies and field amplitudes. In Fig.~\ref{fig:rabi}(a,d), the field contains a resonant component at $\omega_1=\Delta=40$ together with a second-harmonic component at $\omega_2=2\Delta=80$, with $E_1=4$ and $E_2=4$. The resonant component produces the dominant coherent population cycling, while the $2\Delta$ component contributes rapid dressing-induced modulations, visible as higher-frequency ripples superimposed on the primary oscillation. In Fig.~\ref{fig:rabi}(b,e), both components are detuned ($E_1=4$, $E_2=4$, $\omega_1=36$, and $\omega_2=44$), leading to strongly reduced net population transfer. Finally, Fig.~\ref{fig:rabi}(c,f) shows that the long-time population exchange is governed primarily by the resonant component: with $E_1=1$ at $\omega_1=\Delta=40$ and a stronger off-resonant second harmonic ($E_2=4$ at $\omega_2=2\Delta=80$), slow Rabi oscillations persist with a timescale set mainly by the weaker resonant field, whereas the stronger high-frequency component mainly adds fast oscillatory dressing rather than dominating the net population exchange.

\subsection{Simple Avoided Crossing}

\begin{figure*}[t]
\centering
\includegraphics[width=\textwidth]{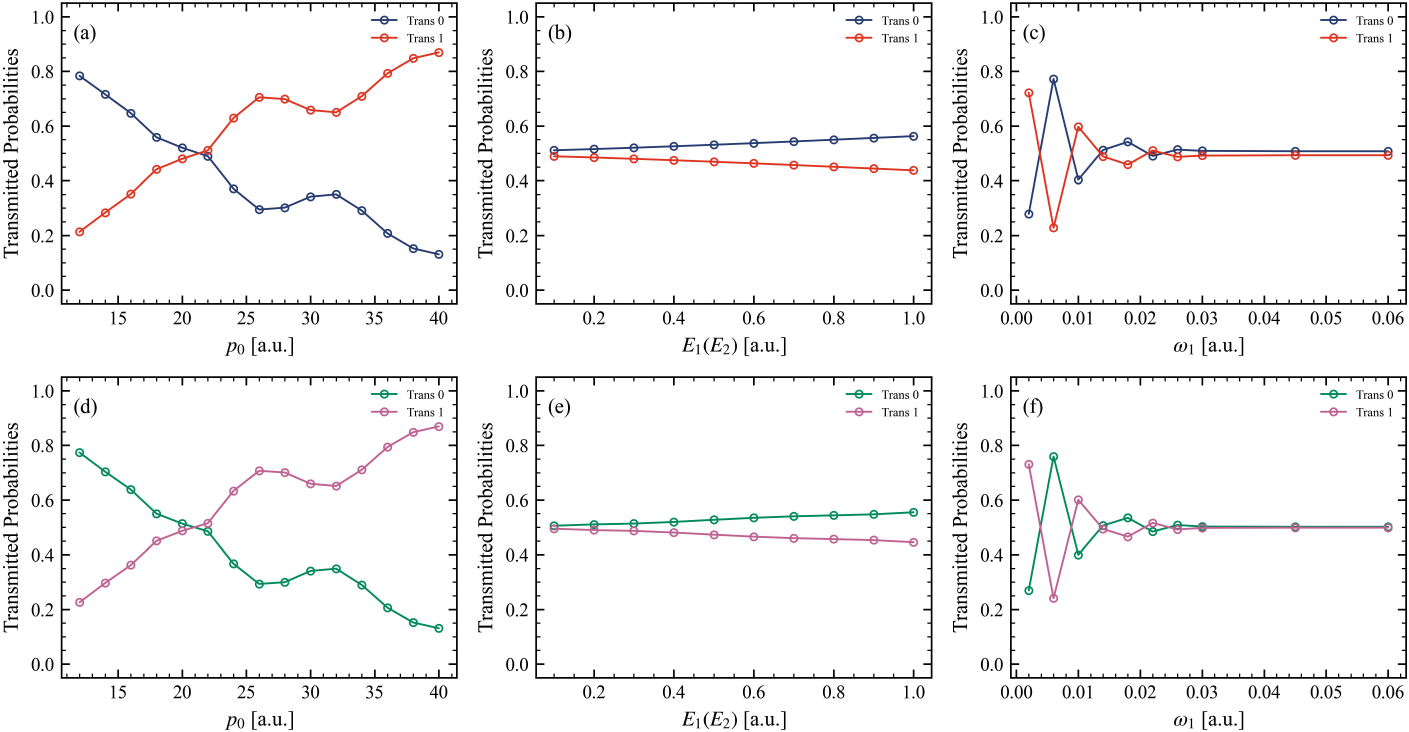}
\caption{Transmitted probabilities on the lower surface (Trans 0) and the upper surface (Trans 1) for the driven simple avoided-crossing model. Top row: numerically exact split-operator results. Bottom row: two-mode F-FSSH results using $10^{4}$ trajectories and a time step $\Delta t=0.5$. Left column: scan over the incident momentum $p_0$ with $E_1=E_2=0.3$, $\omega_1=0.02$, and $\omega_2=2\omega_1$. Middle column: scan over field amplitudes $E_1=E_2$ with $p_0=20$, $\omega_1=0.02$, and $\omega_2=2\omega_1$. Right column: scan over the driving frequency $\omega_1$ (with $\omega_2=2\omega_1$) with $p_0=20$ and $E_1=E_2=0.3$. Overall, two-mode F-FSSH closely matches the split-operator benchmarks for the transmitted probabilities across the momentum, field amplitude, and driving frequency scans.}
\label{fig:tully1}
\end{figure*}

We next consider the driven simple avoided-crossing model, where the two-frequency field modulates the diabatic coupling in the interaction region and thereby controls nonadiabatic scattering. Fig.~\ref{fig:tully1} compares transmitted probabilities from numerically exact split-operator propagation (top row) with those from two-mode F-FSSH (bottom row). For two-mode F-FSSH, we use an ensemble of $10^{4}$ trajectories with a time step $\Delta t=0.5$. Across these scans, two-mode F-FSSH closely tracks the split-operator benchmarks, indicating that it captures the field-controlled scattering dynamics over a wide range of parameters.

The left column [Fig.~\ref{fig:tully1}(a,d)] shows the transmitted probabilities as a function of the incident momentum $p_0$ at fixed driving parameters ($E_1=E_2=0.3$, $\omega_1=0.02$, and $\omega_2=2\omega_1$). As $p_0$ increases, the traversal time through the localized interaction region decreases, reducing the effective action of the time-dependent coupling and leading to a systematic redistribution of transmission between the lower and upper surfaces. The middle column [Fig.~\ref{fig:tully1}(b,e)] scans the field amplitudes $E_1(=E_2)$ at fixed $p_0=20$ and fixed frequencies ($\omega_1=0.02$ and $\omega_2=2\omega_1$). The transmitted probabilities vary smoothly with increasing field strength, consistent with an effective-coupling picture: a larger modulation depth enhances the time-dependent mixing in the interaction region and thus biases the net transfer between the lower and upper surfaces. The right column [Fig.~\ref{fig:tully1}(c,f)] scans the driving frequency $\omega_1$ (with $\omega_2=2\omega_1$) at fixed $p_0=20$ and $E_1=E_2=0.3$. At low frequencies, the transmitted probabilities exhibit pronounced oscillations. These oscillations arise from interference between multiple nonadiabatic transition pathways, which accumulate different dynamical phases while the nuclear wavepacket traverses the time-modulated interaction region. As $\omega_1$ increases, the oscillations are rapidly suppressed and the transmission approaches a nearly frequency-independent limit with $\mathrm{Trans~0}\approx \mathrm{Trans~1}\approx 0.5$. This crossover reflects a high-frequency averaging effect: when the coupling oscillates much faster than the scattering timescale, phase-dependent interference between different transition pathways is effectively averaged out, and the dynamics is governed by an effective time-averaged coupling.

\subsection{Dual Avoided Crossing}

\begin{figure*}[t]
\centering
\includegraphics[width=0.675\textwidth]{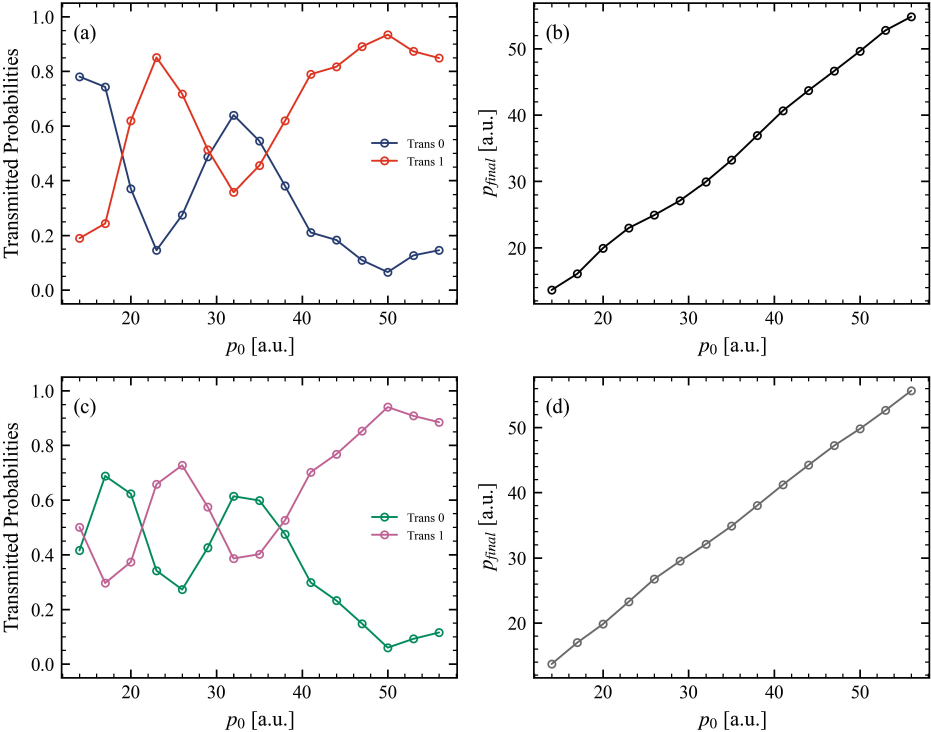}
\caption{Transmitted probabilities on the lower surface (Trans 0) and the upper surface (Trans 1) (left column) and the final nuclear momentum $p_{\mathrm{final}}$ (right column) for the driven dual avoided-crossing model. Top row: numerically exact split-operator results. Bottom row: two-mode F-FSSH results using $10^{4}$ trajectories with $\Delta t=0.5$. Driving parameters: $E_1=E_2=0.3$, $\omega_1=0.05$, and $\omega_2=2\omega_1$. The horizontal axis is the incident momentum $p_0$. Two-mode F-FSSH reproduces $p_{\mathrm{final}}$ nearly quantitatively across the scanned $p_0$ range, while the transmitted probabilities show noticeable deviations at lower $p_0$ but improve at higher $p_0$.}
\label{fig:tully2}
\end{figure*}

We now turn to the driven dual avoided-crossing model. We fix the two-frequency field parameters at $E_1=E_2=0.3$, $\omega_1=0.05$, and $\omega_2=2\omega_1$, and vary the incident momentum $p_0$. Fig.~\ref{fig:tully2} reports the transmitted probabilities together with the final nuclear momentum $p_{\mathrm{final}}$. Compared with the split-operator benchmarks (top row), two-mode F-FSSH reproduces $p_{\mathrm{final}}$ in great agreement across the entire $p_0$ range, whereas the transmitted probabilities show more noticeable deviations at lower $p_0$ and improve progressively as $p_0$ increases.

In Fig.~\ref{fig:tully2}(a,c), the transmitted probabilities exhibit pronounced oscillations as a function of $p_0$, consistent with St\"uckelberg-type interference in a dual avoided-crossing geometry. The two successive nonadiabatic passages generate multiple pathways, and the resulting interference depends on the dynamical phase accumulated during the nuclear propagation between the two crossings. At lower $p_0$, the longer traversal time amplifies the sensitivity of the final populations to phase accumulation and quantum interference, which can lead to an apparent phase shift of the oscillations in independent-trajectory surface hopping. Similar behavior has been reported for this model in Tully's original work.\cite{Tully1990molecular} At higher $p_0$, the nuclear motion is faster and the momentum-resolved transmission becomes less sensitive to such phase effects, which makes two-mode F-FSSH more reliable. Meanwhile, Fig.~\ref{fig:tully2}(b,d) shows that $p_{\mathrm{final}}$ increases approximately monotonically with $p_0$, with small modulations correlated with the electronic outcome. The two-mode F-FSSH results for $p_{\mathrm{final}}$ are essentially indistinguishable from the exact benchmarks.

\section{Conclusions}
\label{sec:Conclusions}

In this work, we developed a two-mode Floquet fewest switches surface hopping (two-mode F-FSSH) approach for two-frequency laser-driven nonadiabatic dynamics. Benchmarking on three driven one-dimensional two-state models, we found that two-mode F-FSSH captures the numerically exact results obtained from the split-operator method over a broad range of field parameters and initial conditions. This includes regimes where two-frequency driving reshapes Rabi-type population transfer, as well as regimes where field-controlled transitions dominate avoided-crossing scattering. Together, these results demonstrate that two-mode F-FSSH is a practical and conceptually transparent route for simulating periodically driven nonadiabatic processes with two-frequency fields.

Despite these promising results, several limitations and open directions remain. First, the present algorithm can be computationally demanding. Although the time-independent Floquet framework can permit larger integration time steps than instantaneous adiabatic fewest switches surface hopping (IA-FSSH) in many situations, convergence with respect to the two-mode truncation often requires sufficiently large $N_1$ and $N_2$. As a result, the two-mode F-FSSH dynamics can become expensive to propagate in practice. Therefore, further efficiency improvements will be important. Possible strategies include a more aggressive separation of nuclear and electronic time steps.\cite{Jain2016an,Zhou2019a,Zhou2020nonadiabatic,Zhou2023nonadiabatic}

Second, extending the present two-mode FSSH framework to pulsed laser fields raises additional conceptual and numerical questions. Floquet theory can render the strictly periodic carrier dynamics time-independent, but slowly varying pulse envelopes reintroduce explicit time dependence.\cite{Bajo2012mixed} A key open issue is how to treat this residual time dependence within a surface hopping formulation. One option is to adopt a slow-envelope (adiabatic) approximation: if the envelope $\mathcal{E}(t)$ varies slowly, then within any single cycle $t\in[t_0,t_0+T)$ one may approximate $\mathcal{E}(t)\approx \mathcal{E}(t_0)$. Alternatively, one may incorporate the envelope-induced time-dependent contributions to the nonadiabatic couplings in the spirit of instantaneous adiabatic treatments. The accuracy and practical trade-offs of these choices remain to be assessed in future work.

Third, the current algorithm is formulated for closed-system coherent dynamics. Extending two-mode F-FSSH to open-system settings, such as nonadiabatic dynamics of molecules coupled to metallic surfaces or electrodes, is an important future direction. Building on our previous one-mode F-FSSH formulations for open systems,\cite{Wang2023nonadiabatic_FSH,Wang2023nonadiabatic_FEF,Wang2024nonadiabatic} such an extension would enable simulations of quantum-transport phenomena under two-frequency driving. One particularly appealing application is to generalize previously demonstrated enhancement of the chiral-induced spin selectivity (CISS) effect using circularly polarized light (CPL).\cite{Liu2025enhancement} Extending to two-frequency driving may offer finer control over the CISS response and broaden the accessible control landscape.

\section*{Acknowledgments}
W.D.\ acknowledges financial support from the National Natural Science Foundation of China (Grant Nos. 22361142829 and 22273075) and the Zhejiang Provincial Natural Science Foundation (Grant No. XHD24B0301). The authors acknowledge computational resources and technical support from the High-Performance Computing Center at Westlake University. J.H.\ thanks Dr.~Yu Wang and Dr.~Yanze Wu for helpful discussions and valuable suggestions.

\section*{Author Declarations}

\subsection*{Conflict of Interest}
The authors have no conflicts to disclose.

\subsection*{Author Contributions}
\noindent\textbf{Jiayue Han:} Methodology (equal); Software (lead); Investigation (lead); Formal analysis (lead); Validation (lead); Data curation (lead); Visualization (lead); Writing -- original draft (lead); Writing -- review \& editing (equal). \textbf{Vahid Mosallanejad:} Methodology (equal); Software (equal); Writing -- review \& editing (equal). \textbf{Ruihao Bi:} Methodology (equal); Software (equal); Investigation (equal); Writing -- original draft (supporting); Writing -- review \& editing (supporting). \textbf{Wenjie Dou:} Conceptualization (lead); Supervision (lead); Funding acquisition (lead); Resources (lead); Project administration (lead); Methodology (equal); Writing -- review \& editing (lead).

\section*{Data Availability}
The data that support the findings of this study are available from the corresponding author upon reasonable request.

\bibliographystyle{aipnum4-1}
\bibliography{references}

\end{document}